\documentclass[12pt,preprint]{aastex}
\usepackage{graphicx}
\shorttitle{Hot Components in Symbiotics} 

\shortauthors{Sion et al.} 

\begin{document}

\title{FUSE Spectroscopy of the Accreting Hot Components in Symbiotic Variables}

\author{Edward M. Sion, Patrick Godon} 
\affil{Department of Astrophysics \& Planetary Science, Villanova University, \\ 
800 Lancaster Avenue, Villanova, PA 19085, USA}
\email{edward.sion@villanova.edu ; patrick.godon@villanova.edu}

\author{Joanna Mikolajewska}
\affil{Copernicus Astronomical Center, \\Warsaw, Poland}
\email{mikolaj@camk.edu.pl}

\author{Bassem Sabra}
\affil{Dept. of Physics \& Astronomy, Notre Dame University, Louaize; \\Zouk Mosbeh, Lebanon}
\email{bsabra@ndu.edu.lb}

\author{Craig Kolobow}
\affil{Department of Physics \& Astronomy, Florida Institute of Technology, \\ Melbourne, FL 98195, USA}
\email{craig.kolobow@villanova.edu}

\begin{abstract}

We have conducted a spectroscopic analysis of the 
far ultraviolet archival spectra of four 
symbiotic variables, EG And, AE Ara, CQ Dra and RW Hya. 
RW Hya and EG And have never had a recorded 
outburst while CQ Dra and AE Ara have outburst histories. We analyze these systems while they are in quiescence in order to help reveal the physical properties of their hot components via comparisons of the observations with optically thick accretion disk models and NLTE model white dwarf photospheres. We have extended the wavelength coverage down to the Lyman Limit with FUSE spectra. We find that the hot component in RW Hya is a low mass white dwarf with a surface temperature of 
160,000K. 
We re-examine whether or not the symbiotic system CQ Dra is a triple system 
with a red giant transferring 
matter to a hot component made up of a cataclysmic variable in which the  
white dwarf has a surface temperature as low as $\sim$20,000K. 
The very small size of the hot component contributing to the shortest 
wavelengths of the FUSE spectrum of CQ Dra agrees with
an optically thick and geometrically thin ($\sim$4\% of the WD surface) 
hot ($\sim 120,000$K) boundary layer. Our analysis of EG And reveals that its hot component is a hot, bare, low mass white dwarf with a surface temperature of 80-95,000K, with a surface gravity $\log(g)= 7.5$.
For AE Ara, we also find that a low gravity ($\log(g) \sim 6$) 
hot ($T \sim 130,000$K) WD accounts for the hot component.
\end{abstract}


\section{Introduction}

Up to now, virtually all of the temperatures and luminosities of 
symbiotic hot components have been derived using the modified Zanstra 
method to obtain radiation temperatures needed to account for 
photoionization of the observed emission lines \citep{mue91,mue94,sko05},  
or by simply assuming a photospheric temperature. This is understandable given 
the daunting complexity of symbiotic stars and the possible sources 
of FUV continuum radiation that may arise from that complexity. 
Nevertheless, many models have been proposed for the origin of the 
emission lines including the assumption that they all arise from 
photoionization. Moreover, the temperatures are derived for the most 
part from the He II (1640) recombination line and the H lines, but 
they yield different temperatures, with the temperature from the 
He II lines much higher than that from the H lines. The He II lines may
arise from a wind or from wind collision shocks or coronae. Added 
to this uncertainty is the fact that for the very few systems analyzed 
up to now with actual NLTE photospheric models (mostly post-novae), 
the model-derived temperatures for hot accreting white dwarfs are 
a factor of two to three lower than the temperatures derived by Zanstra techniques.

An important archive of far ultraviolet spectra on symbiotic stars
was obtained using the Far Ultraviolet Spectroscopic Explorer (FUSE) spacecraft.
These spectra cover a wavelength range from 1180\AA\ 
down to the Lyman Limit at 912\AA.
For observations of the hot accreting components of symbiotic stars, this offers a huge advantage
because, unlike the IUE SWP spectral range, the FUSE spectra are unaffected by any UV contribution of
the nebular continuum which arises from the photoionization of the red giant wind. 
In IUE and HST FOS, GHRS and STIS spectra, this nebular continuum 
tends to flatten their continuum slopes which, if not removed properly, 
leads to grossly underestimated effective temperatures and accretion rates.  

Moreover, in the short wavelength part of the FUV, the hot 
boundary layer (BL) between the star and disk
may contribute to the FUV.  
\citet{lun13} have shown that most of the symbiotic systems 
they observed in the X-ray domain  with the SWIFT spacecraft appear 
to have optically thick boundary layers that emit strongly in the FUV. 
Given this empirical result, it is clear that our modeling of the 
FUSE spectra should, when possible, 
take into account the boundary layer between 
the hot star and disk in symbiotic stars. 

We have selected four systems with archival FUSE spectra for this study: 
CQ Dra, RW Hya, EG And and AE Arae. 
For CQ Dra and RW Hya we found, respectively, 
a matching IUE SWP spectrum and HST/GHRS spectrum  
to extend the spectral coverage.  
There is no evidence that the S-type symbiotics in this paper are metal poor, nor that their giant donor stars are 
s-process enhanced (Galan et al.2016). This s-process enhancement is due to past mass transfer 
(when the present WD was on the AGB), and provides very strong indications that the
WD mass should be at least $\sim 0.55 M_{\odot}$.
CQ Dra has an outburst history and is clearly more interacting than RW Hya. 

For those systems showing ellipsoidal light curves, especially in the red and 
near-IR range (where the red giant dominates the spectrum), it is very likely 
that the red giant is filling or nearly filling its Roche lobe.
Therefore the mass transfer and accretion is mostly through L1, and hence
an accretion disk may be present. Their mass 
transfer/accretion rates should also be relatively high, 
$\sim 10^{-7} M_{\odot}$/yr as revealed by 
numerical hydrodynamic simulations of mass transfer in such systems. In relation to this mass transfer, 
\citet{sha16} have found, using SWIFT X-ray observations, that the wind from the giants in S-type symbiotic stars is not spherically symmetric but is enhanced or focused in the orbital plane which raises the accretion efficiency onto the hot component.

The published orbital and 
physical parameters of these four systems are given in Table 1 where we list, 
by row: 
(1) the system name; 
(2) the distance estimate in pc;
(3) orbital period in days $P_{orb}$; 
(4) the orbital inclination in degrees; 
(5) estimated mass of the hot component, $ M_{hot}$; 
(6) the color excess E(B-V); 
and (7) the references for these parameters. 
These systems have been studied while in quiescence which optimizes the study of their hot components because their accretion rates and system luminosities tend to be lower. 
Our aim is to determine 
for the first time, with NLTE model accretion disks and model photospheres, the 
accretion rates, accretion efficiency and the temperatures and luminosities of the hot 
components and ascertain whether their FUV radiation is from accretion light or WD photospheric light or both. 

In Sec.2 we present an observing log of the archival FUV spectra 
utilized in our study;
in Sec.3 we describe our model accretion disks, high gravity photospheres 
and modeling analysis techniques; 
in Sec.4 we describe each system in greater depth and summarize our model 
fitting results; and in Sec.5 we draw our conclusions.

\section{Archival Far Ultraviolet Spectroscopic Observations}

In Table 2, we provide the observing log for all four systems where we list the 
observing details by row: 
(1) the system name; 
(2) the spectrum data identification number ; 
the (3) date and (4) time of the observation; 
(5) the exposure time in seconds.  

All of the FUSE spectra of the four systems were acquired through the LWRS aperture in TIME-TAG mode. Each exposure was made up of multiple subexposures. 
As pointed out by \citet{god12}, the FUSE reduction requires extensive
post-processing of the data. Details on the acquisition and processing of 
the FUSE data as well as the potential pitfalls (e.g. the `worm', fixed-pattern
noise, etc..) are discusssed by \citet{god12} and will not be repeated here. 
The co-added FUSE spectra of EG And, AE Arae, CQ Dra, and RW Hya are displayed in Figs. 1,2,3 and 4,respectively, where the strongest spectra line features are identified. 

We also used an IUE SWP spectrum to extend the model fits to the FUSE 
spectrum of CQ Dra to $\sim $2000\AA\ , and an HST/GHRS spectrum 
to extend the model fits to the FUSE spectrum of RW Hya to $\sim$1800\AA\ ,
as listed in Table 2.  

\section{Model Accretion Disks and Model Photospheres}

Model accretion disks from the optically thick disk model 
grid of \citet{wad98} were utilized in this study. The outermost disk radius, $R_{out}$ is chosen in such a way that the $T_{eff}(R_{out})$ of the outermost annulus is close to 10,000K. Disk annuli beyond this are so cool that 
they would contribute very little to the mid and far UV disk flux. For these steady state disks, by definition the mass transfer is assumed constant for all annuli.

Hence, the disk temperature as a function of radius is given by

\begin{equation}
T_{eff}(r)= T_{s}x^{-3/4} (1 - x^{-1/2})^{1/4}
\end{equation}

where  $x = r/R_{wd}$
and $\sigma T_{s}^{4} =  3 G M_{wd}\dot{M}/8\pi R_{wd}^{3}$

These disk models include limb darkening which is incorporated using the methodology of \citet{dia96}.
They use the Eddington-Barbier relation,the kinetic temperature increase with depth in the disk, and the temperature and wavelength dependence of the Planck function. 

The database of disk models has the following combination 
of disk inclination angle i = 18, 41, 60, 75, 81 degrees, 
accreting WD mass $M_{wd} $= 0.35, 0.55, 0.80, 1.03, and 1.21 $M_{\odot}$ and 
mass accretion rate $Log(\dot{M}) $= -8.0, -8.5, -9.0, -9.5, -10.0, -10.5 
to fit the observed spectrum. When called for, we also constructed model disks outside of the parameter range of \citet{wad98} in our search for the best fitting models. 

Theoretical, high 
gravity, solar composition photospheric spectra were computed by 
first using the code TLUSTY version 200 \citet{hub88} to calculate 
the atmospheric structure and then SYNSPEC version 48 \citep{hub95}  
to construct synthetic spectra. We used our database of photospheric 
spectra covering the temperature range from $\sim$13,000K to 250,000K in 
increments of 500K (for low $T_{wd}$), 1000K (for intermediate $T_{wd}$), 
5000K and 10,000K (for very large $T_{wd}$), and a surface gravity range, 
$\log(g) $= 7.0 - 9.0, in increments of 0.2 and 0.5 in $\log(g)$. When needed, we have also computed models outside this range of surface gravities in our search for better fitting solutions. The mass radius relation
(from \citet{ham61,woo95} or \citet{pan20} for different compositions
and non-zero temperature WDs) is used to obtain the radius of the WD.
When needed, we also included a boundary layer by assuming that a fraction
of the stellar surface (the equatorial region) has a higher temperature. 
Namely, we generated a two-temperature WD spectrum to include the boundary 
layer. The size and the temperature of the boundary layer were varied 
until a best-fit was found.  
  
When available, we adopted published E(B-V) values to de-redden the spectra.  
We took the published parameters for the four systems 
to use as initial estimates in our model fitting. 
The emission line and ISM absorption line regions  
were masked out in the model fitting. 
We carried out a range of disk and photosphere model 
fits to derive accretion rates, possibly estimate hot component 
temperatures, masses, orbital 
inclinations, and to compare model-derived distances with distance estimates 
published in the literature. Our main focus is to determine whether an accretion 
disk is present, estimate the temperature of the white dwarf, estimate the 
accretion rate, and gain possible insights into the underlying mechanism of
their outbursts. The underlying, accreting white dwarf could be hidden from 
us by the accretion disk (if a disk can form in a wind-accreting system) or 
by the nebula and red giant's wind. It is because of the multi-component flux 
contributions of these emitting sources that actual model fitting analyses of 
the hot components have been very few and daunting.  
 For both of these two types of models, we determined  
whether an improved fit resulted from a combination of a disk plus a white dwarf model. We employ a chi square minimization routine FIT, to find the lowest reduced $\chi^{2}_{\nu}$ ($\chi^2$ per degrees of 
freedom $\nu$) values model (best-fit).   
The scale 
factor, $S$, normalized to a kiloparsec and solar radius, is defined in terms of the 
white dwarf radius R as

 $F_{\lambda(obs)} = S H_{\lambda(model)}$,
 
where $S=4\pi R^2 d^{-2}$, and $d$ is the  distance to the source. For the range of white dwarf masses, we adopted the mass-radius relation from the evolutionary model 
grid of \citet{woo95} for C-O cores. The best-fitting model or combination of 
models were selected on the basis of the minimum $\chi^{2}_{\nu}$ value achieved. However, we required
additional fitting criteria: the goodness of fit to the continuum slope, the goodness of fit to the 
observed Lyman series (when not contaminated with emission) 
and the requirement that the scale factor-derived 
distance agrees with published distance estimates in Table 1.

\section{System by System Discussion and Modeling Results}

The best fitting results are recapitulated in Table 3 
where for each system, by row, 
we list the log of the WD surface gravity, 
the WD effective surface temperature, the WD radius in solar radii, the WD luminosity in solar units,
the boundary layer temperature (if the BL was included in the best fit model), the boundary layer luminosity (if the 
BL was included in  the best fit model), 
and the mass accretion rate of the accretion disk (if a disk was included in the best fit model).

\subsection{CQ Dra}

CQ Dra was originally thought to be a cataclysmic variable in binary orbit 
with a red giant \citep{rei88}, but was later re-classified as a symbiotic variable. 
However, as discussed below, the cataclysmic variable interpretation of the hot 
component may be correct after all. 
CQ Dra has a reliable HIPPARCOS parallax $\pi=5.25 \pm 0.48 $mas 
\citep{van07}. The orbital period for CQ Dra was determined to be 1703 days by \citet{egg89} from radial velocity 
periodicity searches.
This is the longest orbital period of the four systems discussed in this work. 

We found a usable FUSE spectrum of CQ Dra in the MAST 
archive driving our investigation in the following direction. 
The FUSE spectrum has much lower continuum flux level than the IUE
spectra and was obtained in a state of relative low luminosity. Curiously, the Lyman (series) profiles of H are not clearly seen, making it more difficult to identify  a disk versus a WD. The spectrum does not show signs of interstellar absorption 
as seen in the FUSE spectra of the other 3 systems in this work, 
except for atomic H (sharp Lyman series towards the 
shortest wavelengths). Some of the sharp emission lines are probably 
due to terrestrial airglow and sunlight reflected in the telescope,
and some of the sharp, weak absorptions 
may also be either terrestrial (e.g. N\,{\sc i} 1135) 
or interstellar (e.g. Fe\,{\sc ii} 1145). 

While identifying lines in the FUSE range via comparison with FUSE 
spectra of other systems with accreting white dwarfs, we found that 
the FUSE spectrum of the hot white dwarf in the dwarf nova RU Peg 
during quiescence, is very similar to the FUSE spectrum of CQ Dra,  
as clearly shown in Fig.5. 
There is also a similarity between the IUE spectra (e.g. SWP28355) 
of the two systems as displayed in Fig. 6. At first glance, these similarities would seem to add support the possibility that the hot component in CQ Dra may indeed be a cataclysmic variable 
as \citet{rei88} originally claimed, thus
identifying CQ Dra as a triple system.

For the modeling of the FUSE spectrum, we assume a WD mass of $0.8M_{\odot}$.    
Some parts of the FUSE spectrum show better agreement 
with a WD at $T_{wd}=30,000$K while at the shortest wavelengths in the 
FUSE spectrum, there is better agreement with a WD surface temperature 
higher than 50,000K. 
However, such a high temperature lead to an unacceptably large distance. 
When we fit the FUSE spectrum with an accretion disk 
model, then for a distance of 100 pc (the \citet{rei88} distance), 
we obtain $\dot{M} = 10^{-9.5}M_{\odot}$/yr for disk inclinations of 40 to 
60 degrees. If $\dot{M} = 10^{-9}M_{\odot}$/yr, the distance is of order 250 pc 
and the fit is poor. The only decent accretion disk fit to the FUSE 
spectrum of CQ Dra requires $\dot{M} = 1 \times 10^{-8}M_{\odot}$/yr 
and corresponds to a distance of  $\sim$950 pc. 

Hence the very hot component of the system contributing flux to the 
shortest wavelengths of FUSE must have a very small size, smaller
than the inner disk and smaller than the WD. 
We, therefore, decide to include the contribution of 
a boundary layer in our modeling. 

The BL is included either as a direct optically thick
component (at high mass accretion rate the boundary layer forms
a hot equatorial spread layer on the WD surface \citep{ino99,pir04}),  
or as an optically thin hot component (at low mass accretion rate
\citep{pop93}) 
heating up the equatorial region of the WD through advection of
energy \citep{abr95}.  
Consequently, in either cases, 
we assume, that the WD equatorial region has an 
elevated temperature,
and we model the WD as a two-temperature component. The WD itself has a 
temperature that is moderate ($T < 30,000$ K, so as to not contribute
all of the flux) and the boundary layer has a temperature of the 
order of 100,000 K. If the visible part of the BL has a fractional
area $f$, then the visible part of the WD has an area $1-f$. 
To this two-temperature WD we add an accretion disk. 

We obtain that the best fit to the continuum  and within a reasonable
(scaled) distance is for a WD with a temperature 
$T_{\rm wd} \approx 20,000 \pm 3000$ K, a boundary layer 
of size $4 \pm 1$\% and temperature $120,000 \pm 20,000$ K, 
and a disk with 
$\dot{M} \approx 10^{-10}M_{\odot}$/yr, $i \approx 40-50^{\circ}$.  
Most of the solutions were obtained for 
$\dot{M} = 10^{-10}M_{\odot}$/yr, but some solutions were also
obtained for   
$\dot{M} = 10^{-10.5}M_{\odot}$/yr
and $\dot{M} = 10^{-9.5}M_{\odot}$/yr, therefore adding 
an error bar to the mass accretion rate 
$\log(\dot{M}) = -10.0 \pm 0.5 $ ($< M_{\odot}$/yr $>$). 
We present  such a best fit to the FUSE spectrum of CQ Dra in Fig.7. 
We note that such a mass accretion rate is consistent with the 
low brightness state during which the FUSE spectrum was obtained. 

We also retrieved an IUE SWP spectrum (33521) of CQ Dra in the low state
with a continuum flux matching the FUSE spectrum (in the spectral
region where they overlap), and decided to model the combined 
FUSE + IUE spectrum. We found that the IUE spectrum agrees well 
with our FUSE best-fit model, as shown in Fig.8, an indication that
CQ Dra consistently comes back into the same low state.

\subsection{RW Hya}

The S-type symbiotic system RW Hydrae (=HD 117970) has a primary component classified as a M2 III red giant. The The inclination of RW Hya is high, just sufficient for the system  to undergo eclipses while the orbital period is
370.2 days \citep{mer50,ken95}. Published estimates of the white dwarf mass range extend from 0.3 to 0.6 
$M_{\odot}$. The red giant mass is between 0.5 and 2$M_{\odot}$ 
\citep{sch96,ken95}. We take the distance to the system, 
based upon the corotation of the red giant as found by 
\citet{sch96} of 670 +/- 100 pc. The total luminosity of the system at this distance is
$\sim  700 L_{\odot}$. RW Hya exhibits ellipsoidal variability 
\citep{rut07} based upon the observed light curve. 
The distance set by the Roche lobe geometry, which requires a filling factor above 0.9, is then 1.7 kpc. 

A noteworthy aspect of RW Hya for studies of the hot component is
is that there is little or no evidence for an accretion From the
disk. Its multi-wavelength data was analyzed most recently by \citet{sko05} who dismissed an 
accretion disk but fit the FUV data with a hot black body SED at $T_{eff} = 10^5$ K. 

To our delight, we found a good FUSE spectrum of RW Hya in the MAST 
archive which had never been analyzed and allowed us to sample 
the FUV flux down to the Lyman Limit. The FUSE spectrum of RW Hya,  
with its flat continuum increasing towards shorter wavelengths, 
indicates that the hot component has a very high temperature, 
but the Lyman series is 
affected by ISM atomic H absorption lines. 

Single disk models require a very high mass accretion rate to fit the
relatively flat continuum and yield a ridiculously short distance.
We therefore conclude that the disk does not significantly contribute
to the FUV spectrum, and that the surface area of the hot component
is much smaller than that of an accretion disk. 

We carried out single temperature hot WD model fits for a low mass 
WD to the FUSE spectrum of RW Hya and found a best fit corresponding 
to a white dwarf with a surface temperature of 160,000K with 
$log(g)=6.5$. 
The temperature and gravity are dictated by the shape of the wings of
the Lyman series (where possible) and continuum.  
Assuming a mass of $0.4M_{\odot}$ and radius of $0.065R_{\odot}$
yields a scale factor-derived distance of 811 pc confirming that a low mass 
white dwarf with its larger radius, and a very high temperature gives 
a distance to RW Hya which is within
the error bars of the original Schild et al. distance.
This best-fitting NLTE WD model atmosphere fit to the FUSE spectrum is shown in Fig.9. While we calculate a luminosity of 2 to 3,000$L_{\odot}$ 
as listed in Table 3, a luminosity of 700$L_{\odot}$  
results if one assumes a log(g)=6.8-7.0, instead of 6.5. 
 
While a low mass (large radius) hot component 
seems consistent, it cannot be ruled out that the WD is undergoing 
hydrogen shell burning, has a higher mass but an extended atmosphere 
whose radius is larger than that given by the WD mass-radius relation.

We found a HST GHRS spectrum of RW Hya in the MAST archive that has 
a similar flux level to the FUSE spectrum.
In this way, we extended the wavelength coverage from 1700\AA\  
down to the Lyman Limit. 
After dereddening the spectra with E(B-V)=0.10, we found 
basically the same solution 
for the combined spectrum (FUSE+GHRS) that we found for FUSE alone, namely 
a very hot and very small mass WD can provide the flux needed. 
This single temperature white dwarf fit  is displayed in Fig.10. 
When compared to the model, the HST GHRS spectrum has a flux 
slightly lower. 

An accreting white dwarf at this very high temperature almost certainly sustains thermonuclear burning of the accreted hydrogen and should be a supersoft X-ray source 
The absorbing column of the ionized red giant wind in a symbiotic system is nominally more than sufficient to obscure X-ray emission from an accreting white dwarf with shell burning. Thus, the absence of X-ray emission 
does not rule out the presence of thermonuclear shell burning in the hot component.

\subsection{EG And}

Observations of EG And suggest the wind from the giant 
is focused toward the orbital plane thus raising the accretion efficiency
and rate of accretion onto the WD \citep{sha16}.           
The mass loss rate from the giant is estimated to be
a few times $10^{-6}M_{\odot}$/yr with an inclination of 
i = 80 degrees. The interstellar reddening is E(B-V) = 0.05 \citep{ken16}.  
The temperature ($\sim 75,000$K) and
luminosity of the hot component in EG And were estimated using a modified 
Zanstra method while the orbital period, $P_{\rm orb} = 482.6$ days was 
re-determined by \citet{fek92} who also determined a distance 
of 568 pc. \citet{van07} presented a distance of 
513 $\pm$169 pc based upon a Hipparcos parallax. 
The mass of the hot component $M_{wd} = 0.35 \pm 0.1 M_{\odot}$  
\citet{ken16}. 

We carried out a synthetic spectral analysis of the FUSE spectrum alone. 
We asssumed a distance of 400-700pc, a low white dwarf mass between 
0.35 and 0.6 $M_{\odot}$ and orbital inclination of i = 60, 75,  
and 81 degrees, and an interstellar reddening value of E(B-V) = 0.05.
Our first fitting attempt involved a steady state, optically thick accretion disk. However, our accretion disk fitting met with limited success.The disk fits are poor toward the short wavelength (Lyman Limit) end of the FUSE wavelength range.
Moreover, the distance implied by the disk fits, the scale-factor 
derived distance is far too close, even for 
$\dot{M} = 10^{-8} M_{\odot}$/yr, i = 60 degrees and 75 degrees. 
At a high mass accretion rate, the disk becomes rather thick
and with an inclination of 80 degrees, we expect the boundary layer
near the equatorial region of the star to be masked by the disk. 
We are thus forced to conclude that a standard accretion disk cannot account for the FUSE spectrum of EG And. 

Next, we tried fitting hot NLTE white dwarf photospheres to the FUSE 
data of EG And. The best fit that we obtained is for Log(g)=7.5 and 
$T_{\rm wd} \sim$80-95,000 K. If $\log(g)< 7.5$, then the best-fitting 
temperature becomes lower. 
When we combine an optically thick disk with the WD model, 
there is an improvement over a disk-only fit but the WD +disk fit 
is clearly inferior to the WD-only model. 
The addition of a hot boundary layer to the WD (i.e. a two-tempearture
WD) introduces a degeneracy to the solution in that the contribution of 
small surface area BL with $T\sim 10^5$K BL does not produce a noticeable  
change in the spectrum given that the WD itself has already a 
temperature close to 100,000K (producing a flat spectrum). 
Also because of the high inclination, the contribution from a BL
would be minimal.  
Thus, for EG And, it appears that the hot component can be identified 
with a hot, bare white dwarf having a surface temperature of 
80-95,000K and Log g = 7.5. These derived
parameters using synthetic spectral fitting are in good agreement
with the results of \citet{ken16}. This best-fitting solution is 
displayed in Fig.11. 

\subsection{AE Arae}

On the basis of IR spectroscopy, \citet{fek10} 
found an orbital period of 803 days. 
There is no evidence of eclipses in AE Ara. \citet{fek10} 
estimated an orbital inclination of 51 degrees and a distance range of
2.3 to 3.2 kpc. More recently, \citet{gal16} revealed that AE Ara shows metallicity 
closer to solar by $\sim$0.2 dex. The presence of enriched 14N isotope found in AE Ara reveals that the giant donor has gone through the first dredge-up.

If we assume that the WD mass has to be somewhere between 0.35 and 0.6,
and that the distance is estimated to be around 2.3kpc to 3.5kpc, with 
an inclination of 60 deg and reddening of E(B-V)= 0.25, we find that 
an accretion disk cannot contribute to the flux even with 
$\dot{M}=1 \times 10^{-8}M_{\odot}$/yr. 
Such a disk would imply an extremely short distance and would not
produce the increase of flux observed in the short FUSE wavelengths. 

Here too, 
it is apparent that the hot component in AE Arae is not an accretion disk.
In order to generate such a large FUV flux, a WD has to be hot and 
its radius has to be extended (it cannot be the boundary layer neither). 
We found that fitting solutions with $M_{wd}$ as specified above and the
distance as indicated, must have a gravity as low as $\log(g)= 6.0$ or even lower.  

The best fitting results at $\log(g)= 6$ are as follows.  
For a distance d = 2.3kpc (lower limit), 
the WD temperature ranges between 100,000K and 160,000K,
where the lower temperature is obtained for  
$R_{wd}=0.134 R_{\odot}$ ($M_{wd}=0.65M_{\odot}$), 
and the higher temperatuure is obtained for 
$R_{wd}=0.100 R_{\odot}$ ($M_{wd}=0.36M_{\odot}$). 
For a distance d = 3.5kpc, the only fitting solution is for a larger mass, hotter WD with $M_{wd} = 0.62M_{\odot}$, $R_{wd}=0.131R_{\odot}$ and $T_{wd}$=160,000K. 
An intermediate  best fitting solution is displayed in Fig.12. We found that lower gravity WDs fit better than higher gravity WDs. Hence, with the larger radius of lower mass WDs, the calculated luminosity is higher and is consistent when scaling the model using the radius and distance.

\section{Discussion}

The four S-Type symbiotic binaries that we chose for this study were 
selected for their relative "simplicity" compared with symbiotics containing Mira variables
with their high wind mass loss rates, more complex nebulae and 
associated dust. Our approach was to confront their FUV spectra 
with NLTE accretion disk models and hot white dwarf model 
photospheres, thereby probing the nature of the hot component, testing temperatures derived from 
modified Zanstra techniques, and achieving our key objectives; the accretion rate, accretion efficiency,
and the photospheric temperatures of the hot components in S-type systems. These parameters
are crucial for understanding the evolution of symbiotic stars, including whether accretion disks form and
dominate the FUV light, and whether the relationship between the long term time-averaged accretion rate 
and the surface temperature of the accretors via compressional 
heating \citep{sio95,tow03}, 
also holds for symbiotic hot components. 

For most symbiotics, including the "simpler S-Types",
the FUV spectra obtained with IUE (SWP), and HST (FOS,GHRS,STIS) of the hot components cannot be successfully modeled. The chief problem is that the 
observed FUV continuum slopes are 
less steep due to the inclusion of flux contributions principally from the nebular continuum produced by the photoionized red giant 
wind. Hence, the hot components mimic a much cooler white dwarf or 
cooler accretion disk thereby lowering the derived temperature or 
accretion rate, while the true continuum slope of the hot component 
alone is much steeper \citet{sko05}. 
In addition, while 3D numerical simulations 
\citep{moh07,deval14} 
show that the accretion disks in 
symbiotics resemble optically thick, steady state disks, the scale 
of the symbiotic systems is vastly different from cataclysmics, and 
the accretion flows are more complex. 

However, our investigation 
underscores the critical importance of extending the FUV coverage
of accreting white dwarfs in interacting binary systems down to the 
Lyman Limit. This FUSE coverage was key to our understanding of the nature of the hot 
components in CQ Dra, RW Hydrae, EG And and AE Ara.  

A number of studies of symbiotic stars with X-ray observations by 
\citet{lun07,ken09,lun13,nun16} offer 
important insights into our FUSE FUV observations, 
that those systems with hard X-ray emission ($E >$ several keV) reveal the 
presence of an accretion-disk boundary layer. If the boundary layer is optically
thin, most of the radiated energy (equal to half of the accretion luminosity) is emitted in the X-ray domain,
while an optically thick boundary layer
emits most of its radiation in the EUV/FUV domain, with much less emitted 
in X-rays. For symbiotics without shell burning, the hard X-ray emission can provide an estimate of the accretion rate and the mass of the 
accreting WD. \citet{lun13} have also found that systems in which the 
symbiotic phenomenon is powered by accretion alone (as opposed to shell 
burning + accretion) tend to show large-amplitude, stochastic, UV variability.

In the present work on the four symbiotics observed with FUSE, the X-ray observations of 
EG And by \citet{lun13} offer an important comparison with our FUSE analysis.
\citet{lun13} classifies the X-ray spectrum of EG And as Type $\beta$
which is defined as a Soft X-ray source with most of the photons having energy
less than 2.4 keV, the maximum energy detectable with ROSAT.
Our analysis of the FUSE data indicates that EG And's hot componnent is a bare 
accreting white dwarf (no appreciable disk present) which may emit a hot wind.
Such a wind would interact with the cold, slow wind of the giant donor.
A collision of winds from the
white dwarf with those from the red giant \citep{mue91} 
is the scenario which underlies the $\beta$ classification of
the X-ray spectra of symbiotics. 

\section{Acknowledgements}
We thank an anonymous referee for valuable comments.
This work is supported by NASA Grants NNX13AF12G and NNX13AF11G to Villanova University and by undergraduate summer research support by the NASA Delaware Space Grant Consortium. JM has been supported by the Polish NCN grant no.DEC2011/01/B/ST9/06145.

\clearpage 

\begin{deluxetable}{ccccccl}
\tablewidth{0pc}
\tablecaption{
Physical and Orbital Parameters
}
\tablehead{
System  & Distance  &  $P_{orb}$   & Inclination & $ M_{hot}$  & E(B-V) & Ref. \\       
Name    &  $<$pc$>$ &  $<$days$>$  &   $<$deg$>$ & $M_{\odot}$ &        &         
}
\startdata 
EG And  & 400 - 700 & 482.6        & 60           & 0.4 - 0.6  & 0.05   & (1,2,3)      \\ 
AE Arae & 2.3k-3.5k & 812          & 51           & 0.35-0.67  & 0.25   & (4,5,6)      \\ 
CQ Dra  & 178       & 1703         &   -          &    -       & 0.10   & (7,8)        \\
RW Hya  & 820       & 370.2        & $>70$        & 0.5        & 0.10   & (9,10,11)        \\   
\enddata 
\tablenotetext{1} {\citet{mue91} 
} 
\tablenotetext{2} {\citet{fek00}
} 
\tablenotetext{3} {\citet{ken16}
} 
\tablenotetext{4} {\citet{mik97}
} 
\tablenotetext{5} {\citet{mik03} 
} 
\tablenotetext{6} {\citet{fek10}
} 
\tablenotetext{7} {\citet{rei88} 
} 
\tablenotetext{8} {\citet{perr97} 
} 
\tablenotetext{9} {\citet{mer50}
}
\tablenotetext{11}{\citet{ken95}  
} 
\tablenotetext{10} {\citet{sch96}
} 
\end{deluxetable} 

\clearpage 

\begin{deluxetable}{ccccc}
\tablewidth{0pc}
\tablecaption{
Observation Log	   
} 
\tablehead{
System  & Data ID Obs. &  Date Obs.  & Time (UT)  &  Exp.Time(s) 
}    
\startdata 
EG And  & FUSE C1690190     & 2003-12-01  & 01:20:46   & 9026  \\ 
AE Ara  & FUSE D1460201     & 2004-05-18  & 17:41:56   & 9831  \\ 
CQ Dra  & FUSE D9050401     & 2004-01-09  & 21:24:59   & 36719   \\ 
        & IUE  SWP33521     & 1988-05-13  & 15:55:08   & 1800    \\ 
RW Hya  & FUSE E1460103     & 2004-06-14  & 06:07:21   & 6043  \\          
        & GHRS z2xh0105t    & 1996-03-05  & 18:44:20   &  218  \\          
        & GHRS z2xh0106t    & 1996-03-05  & 18:49:48   &  218  \\          
\enddata 
\end{deluxetable}

\begin{deluxetable}{lcccc}
\tablewidth{0pc}
\tablecaption{
Multi-Component Model Fitting Results
}
\tablehead{
Parameters                & CQ Dra        & RW Hya     & EG And        & AE Ara}
\startdata 
 $\log(g)$ $<$cgs$>$      & 8.4           &  6.5          & 7.5           & 6.0    \\    
 $T_{\rm wd}/10^3$K       & 20$\pm3$      & 160$\pm10$    & 80-95         & 130$\pm30$ \\ 
($R_{\rm wd}/R_{\odot}$)x10$^{-2}$ & 0.94 & 6.6           & 1.9-2.3       & 10-13.4 \\ 
$L_{\rm wd}/L_{\odot}$    & (6.6-22)x10$^{-3}$& (2-3.3)x10$^3$& 12.9-38.4 & (1.6-10.1)x10$^3$ \\ 
 $T_{\rm BL}/10^3$K       & 120$\pm20$    &  ---          & ---           & ---    \\  
$L_{\rm BL}/L_{\odot}$    & 0.316-1.215   &  ---          & ---           & ---        \\ 
$\log(\dot{M}/M_{\odot}$/yr) &$-10\pm0.5$ &  ---          & ---           & ---      \\ 
\enddata 
\end{deluxetable}

\begin{figure}[!ht]
\vspace{-5.cm} 
\plotone{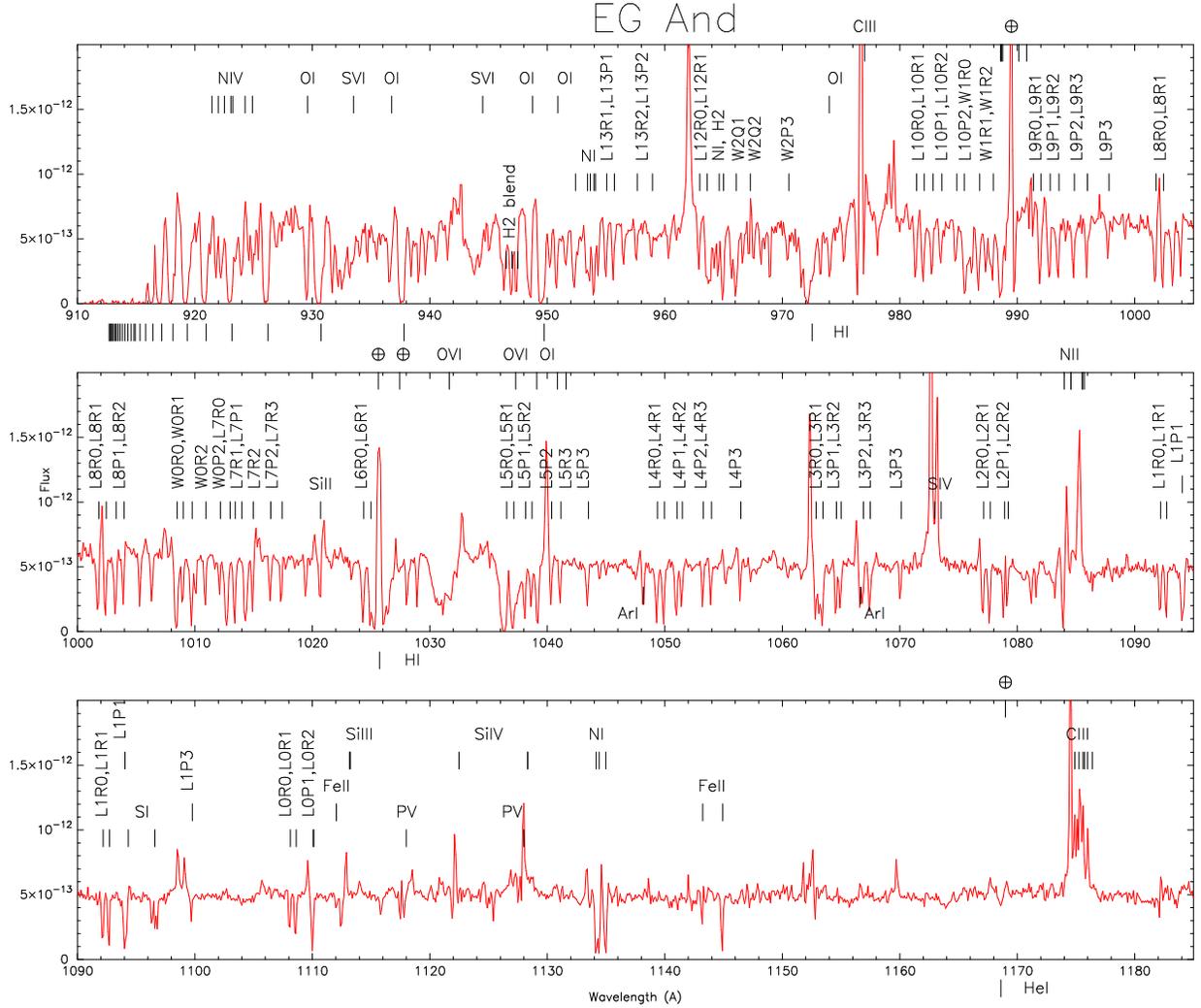}
\vspace{-1.cm} 
\caption{
The FUSE spectrum (flux versus wavelength) of EG And is shown 
with line identification. 
The spectrum has not been dereddened. 
All the sharp absorption lines are possibly
from the interstellar medium (ISM). ISM hydrogen molecular lines are
identified by their band (Werner or Lyman), upper vibrational level 
(here 1 to 13),
and rotational transition (R, P,or Q with lower rotational state J=1-3).
The (ISM) hydrogen Lyman series is identified under each panel. 
Additional ISM lines, such as O\,{\sc i}, Fe\,{\sc ii}, Ar\,{\sc i}
and so on, are also identified.  
Some broad absorption lines from the sources are present, 
and consist of highly ionized species. These are the  
N\,{\sc iv} ($\sim 923$), S\,{\sc vi} ($\sim$933 \& 944), \& O\,{\sc vi}
($\sim$1032 \& 1037) lines. 
Some of the sharp emission lines are also from the sources, such as 
the C\,{\sc iii} (977 and 1175), S\,{\sc iv} (1073) lines and 
the two P\,{\sc v} (1118 \& 1128) lines. 
Other sharp emissions lines are known air-glow lines and are marked 
with an ``earth'' symbol. The nitrogen lines ($\sim$1185 \& 1135) 
are also terrestrial in origin.   
The feature around 1152\AA\ is a known instrument fixed pattern noise (FPN).
}  
\end{figure}

\begin{figure}[!ht]
\vspace{-5.cm} 
\plotone{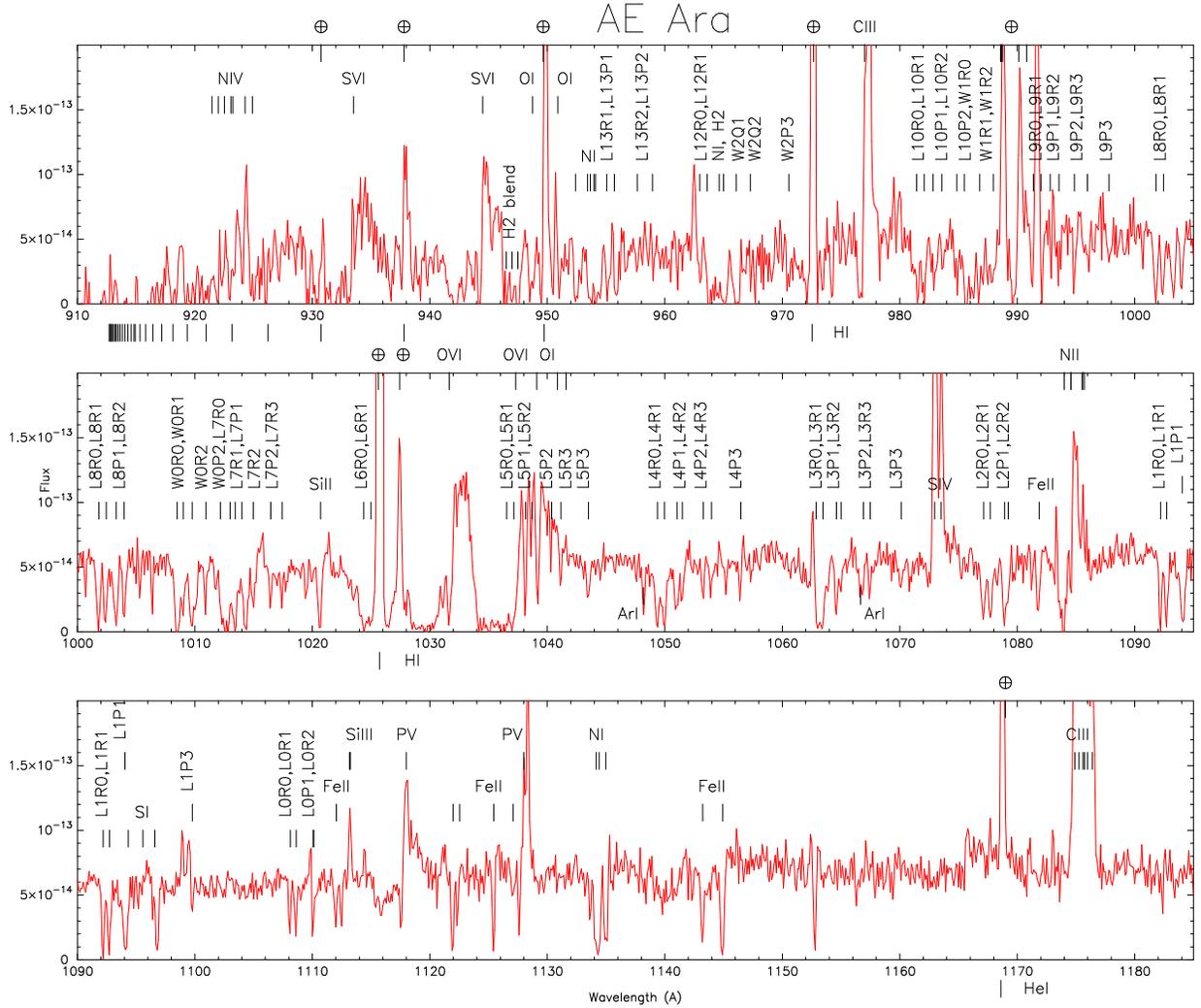}
\vspace{-1.cm} 
\caption{
The FUSE spectrum of AE Ara is shown with line identification. 
The spectrum has not been dereddened. 
Like the spectrum of EG And,
the spectrum of AE Ara is heavily affected by ISM absorption lines, including 
molecular and atomic hydrogen lines, and low ionization species such as  
Fe\,{\sc ii}, Ar\,{\sc i}, O\,{\sc i}, S\,{\sc i}, \& Si\,{\sc ii}. 
The source itself present blue-shifted 
broad absorption lines from higher ionization
species such as N\,{\sc iv} (923), S\,{\sc vi} (933 \& 943), 
and O\,{\sc vi} (1029 \& 1036) as well as 
emission lines from S\,{\sc vi} (934 \& 945), 
C\,{\sc iii} (977 \& 1075), the O\,{\sc vi} doublet, 
S\,{\sc iv} (1073), Si\,{\sc iii} (113),  
and P\,{\sc v} (1118 \& 1128). 
The very sharp emission lines of the
hydrogen Lyman series and O I are most probably geo- and
helio-coronal in origin, 
contamination of sunlight reflected in the SiC channels
(e.g., C III 977). We note that the S VI (933, 944) and the Oxygen
doublet appear to have two components: the first is in emission
at about the rest wavelength, and the second is in absorption
with a blueshift of at least 2\AA\ . 
}
\end{figure}
  
\begin{figure}[!ht]
\vspace{-5.cm} 
\plotone{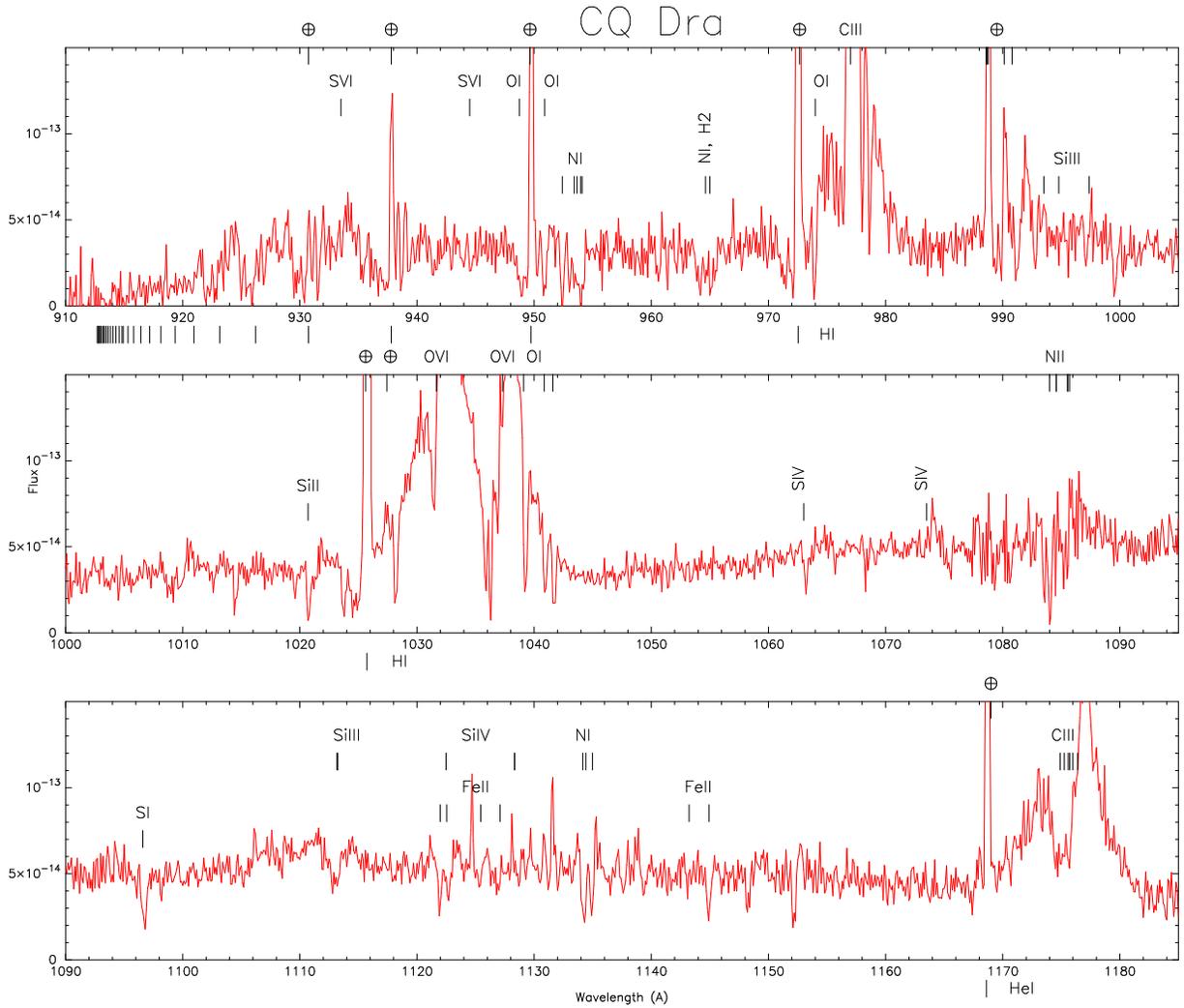}
\caption{
The FUSE spectrum of CQ Dra is shown with line identification.
The spectrum has not been dereddened. 
The spectrum presents little ISM contamination, except possibly for 
some absorption lines from low ionization species such as Fe\,{\sc ii}. 
The source presents broad emission lines from the O\,{\sc vi} doublet 
and the two C\.{\sc iii} lines (977 \& 1175) all superposed with sharper
absorption lines.   
} 
\end{figure}

\begin{figure}[!ht]
\vspace{-5.cm} 
\plotone{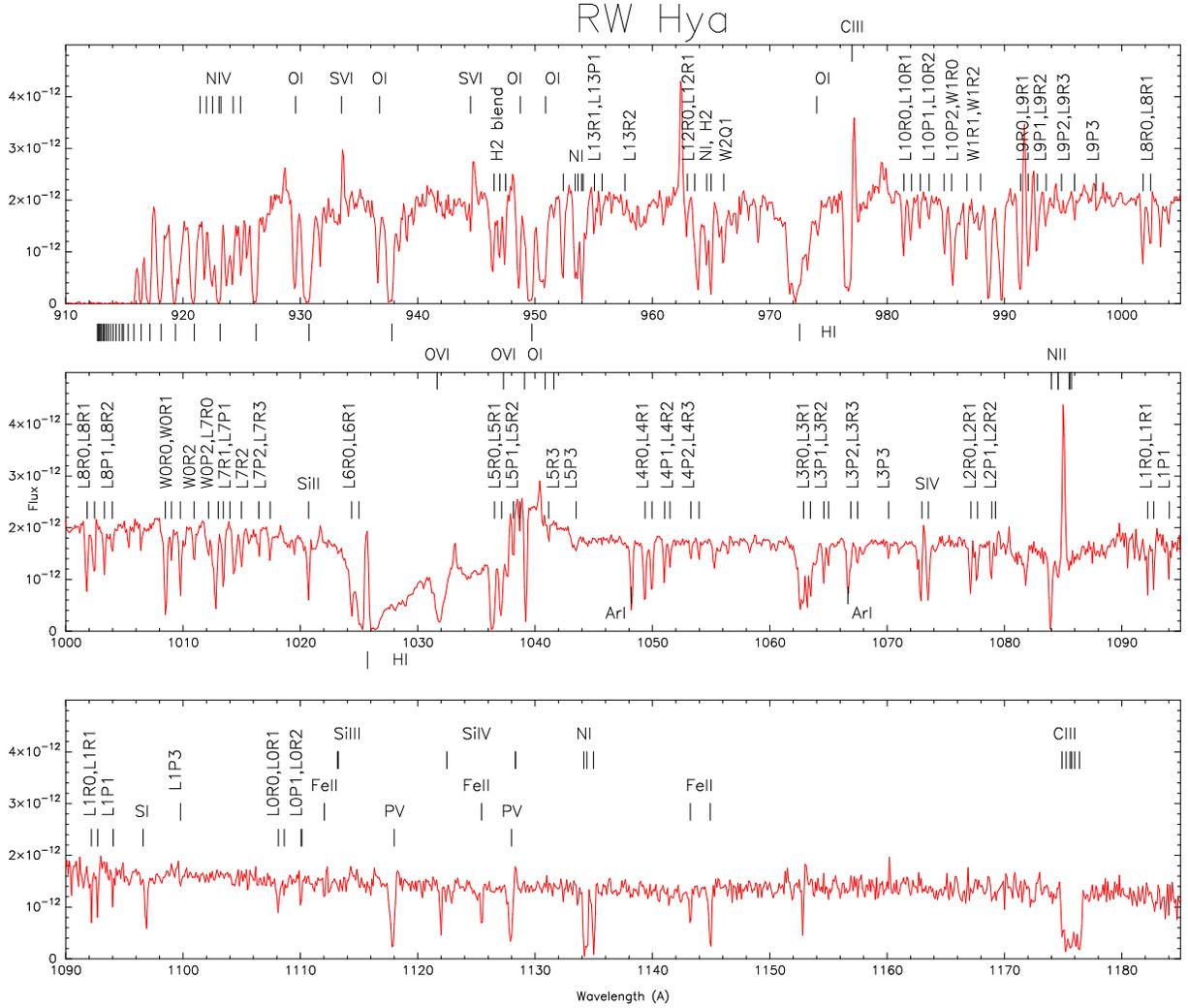}
\caption{
The FUSE spectrum of RW Hya is shown with line identification.
The spectrum has not been dereddened. 
The spectrum is moderately affected by ISM lines. 
The absorption lines we identify from the 
source are  the O\,{\sc vi} doublet lines, the P\,{\sc v} (1118 \& 1128) lines
and the C\,{\sc iii} (1175) line. The O\,{\sc vi} doublet seems to 
be made of a sharp (a few \AA ) component superposed onto a much broader
($\sim$10\AA ) component.  
Weak sharp red-shifted emission lines are identified from higher ionization
species as follows: 
S\,{\sc vi} (934 \& 945), C\,{\sc iii} (977), O\,{\sc vi} (1034),
He\,{\sc ii} (1085) and P\,{\sc v} (1118 \& 1128).   
} 
\end{figure}

\begin{figure}[!ht]
\vspace{-10.cm} 
\plotone{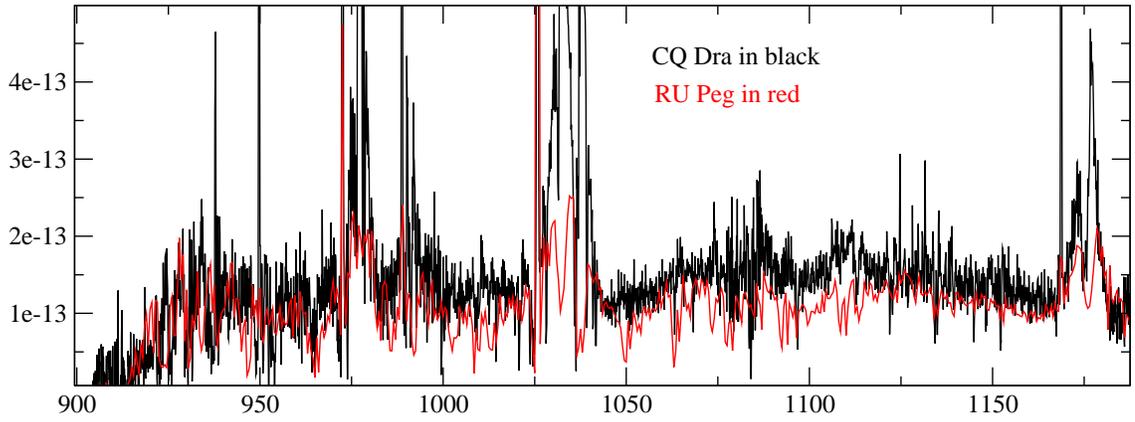}
\caption{The FUSE spectrum of the hot white dwarf in the dwarf nova 
RU Peg during quiescence, shown in comparison with the 
(dereddened) FUSE spectrum of CQ Dra.}
\end{figure}

\begin{figure}[!ht]
\vspace{-10.cm} 
\plotone{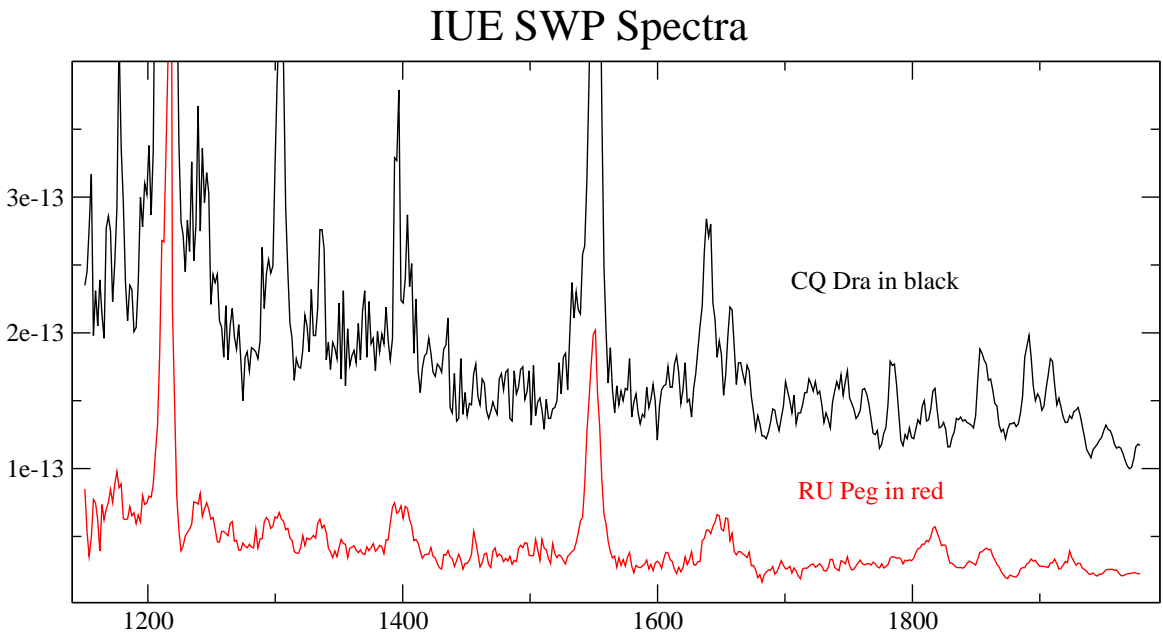}
\caption{A comparison of the IUE spectrum (SWP28355) of CQ Dra
and the white dwarf in the dwarf nova RU Peg during quiescence.
For clarity the IUE spectrum of CQ Dra has not been dereddened here.
The IUE SWP spectra of CQ Dra in the archive are dominated by 
strong emission lines of C III (1175), weak N V (1240) emission, 
weak Si IV (1398) emission, very strong C IV (1550) emission, 
with He II (1640) being notably absent.}
\end{figure}

\begin{figure}[!ht]
\vspace{-5.cm} 
\plotone{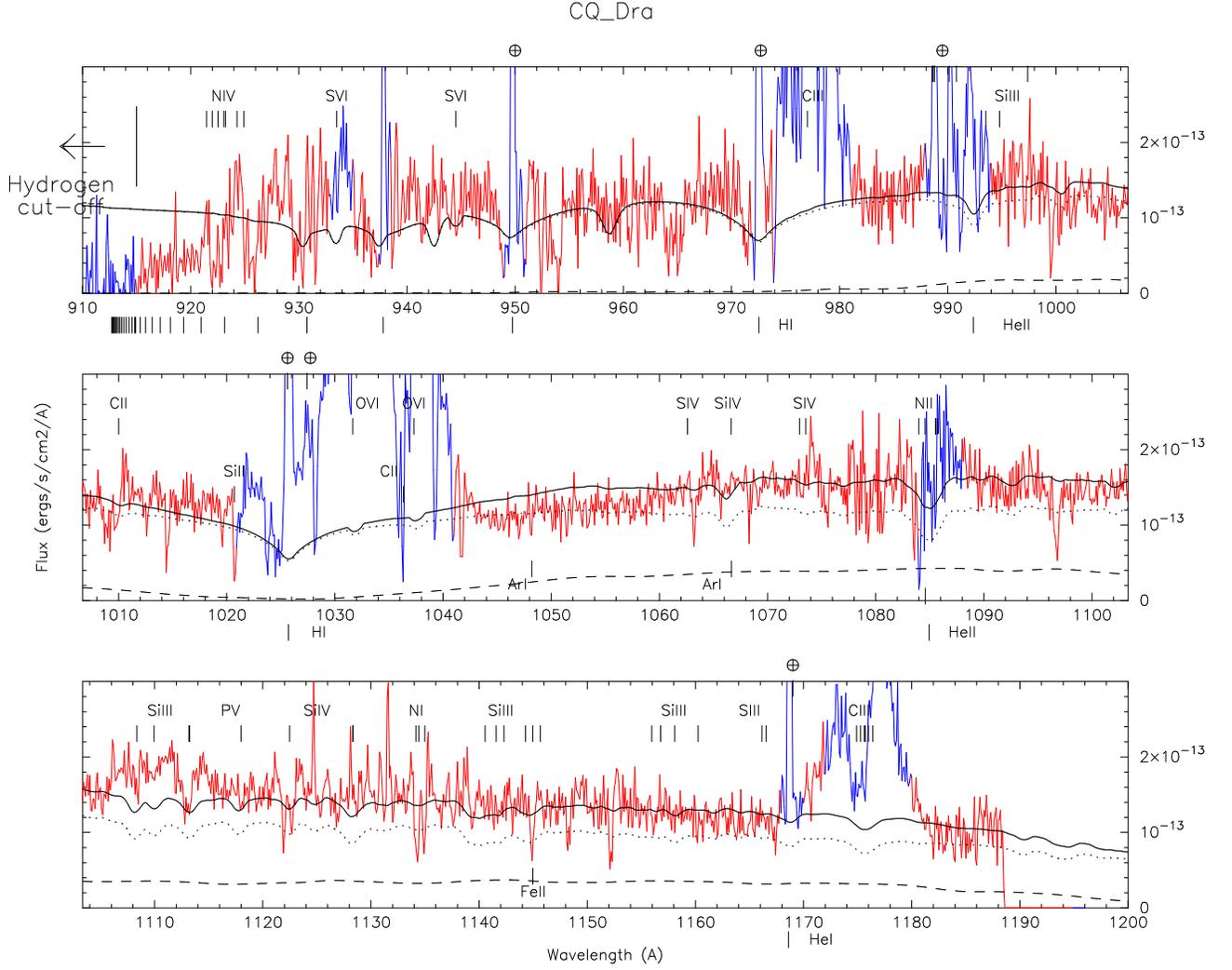}
\caption{The best fit theoretical Disk + white dwarf + boundary layer
spectrum (in black) to the dereddened  
FUSE spectrum of CQ Dra (in red).   
The WD has a mass of $0.8M_{\odot}$ and a relatively low temperature
of 20,000K, with 120,000K boundary layer covering 4\% of the 
WD surface. The disk has an accretion rate of $\dot{M}=10^{-10}M_{\odot}$/yr,
and an inclination of 60$^{\circ}$. The disk (dashed line) 
contributes 18\% of the flux
and the WD+BL (dotted line) 
contribute the remaining 82\%. The distance obtained from
the fit is 164pc.  
The known emission lines 
regions have been masked (in blue) are not modeled. 
} 
\end{figure}

\begin{figure}[!ht]
\vspace{-5.cm} 
\plotone{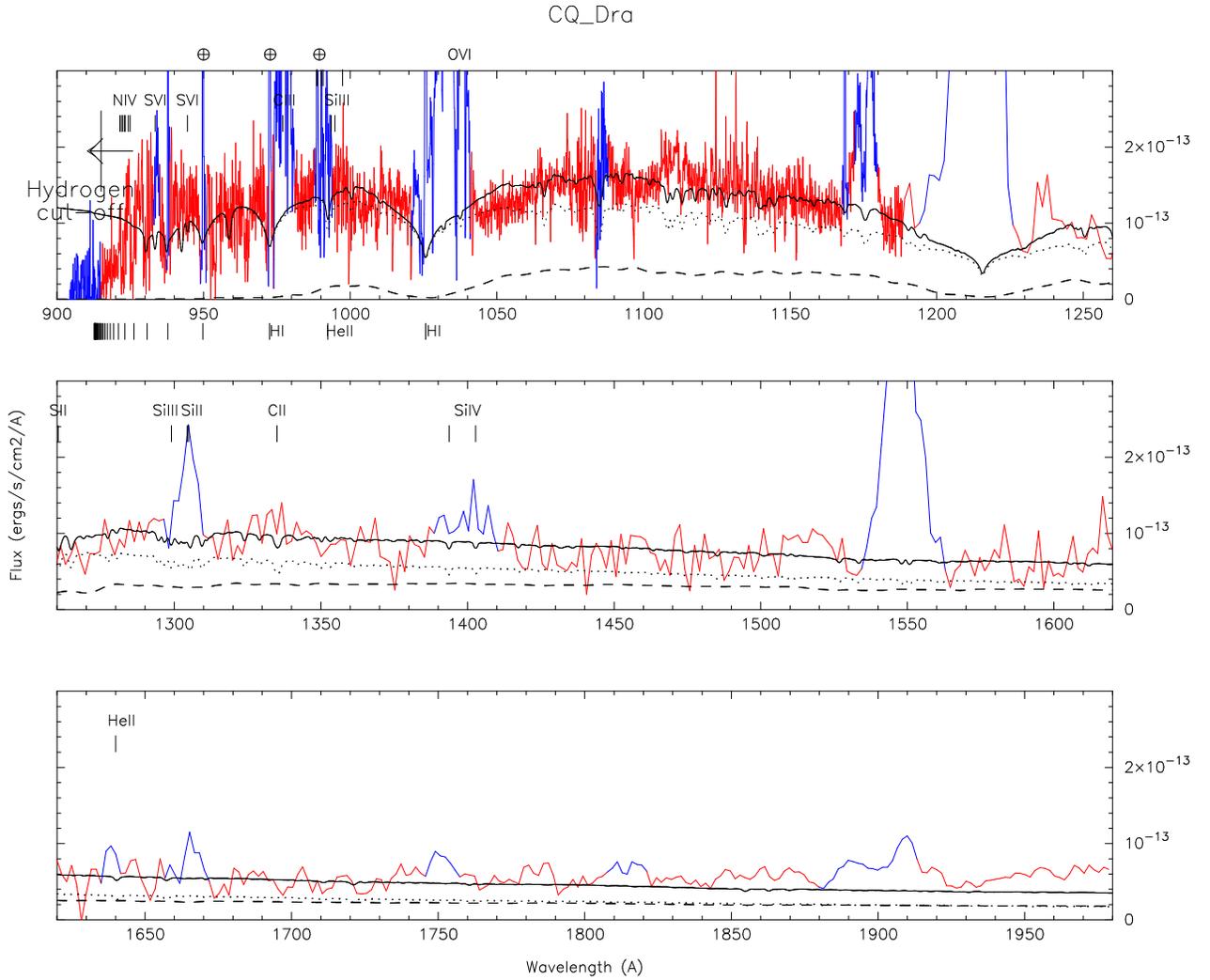}
\caption{
The same model shown in Fig.7 is shown here, but now it fits the 
combined FUSE + IUE SWP (33521) spectrum fo CQ Dra. The fit to the 
continuum is relatively good all the way down to $\sim$1,800\AA\ . 
} 
\end{figure}

\begin{figure}[!ht]
\vspace{-5.cm} 
\plotone{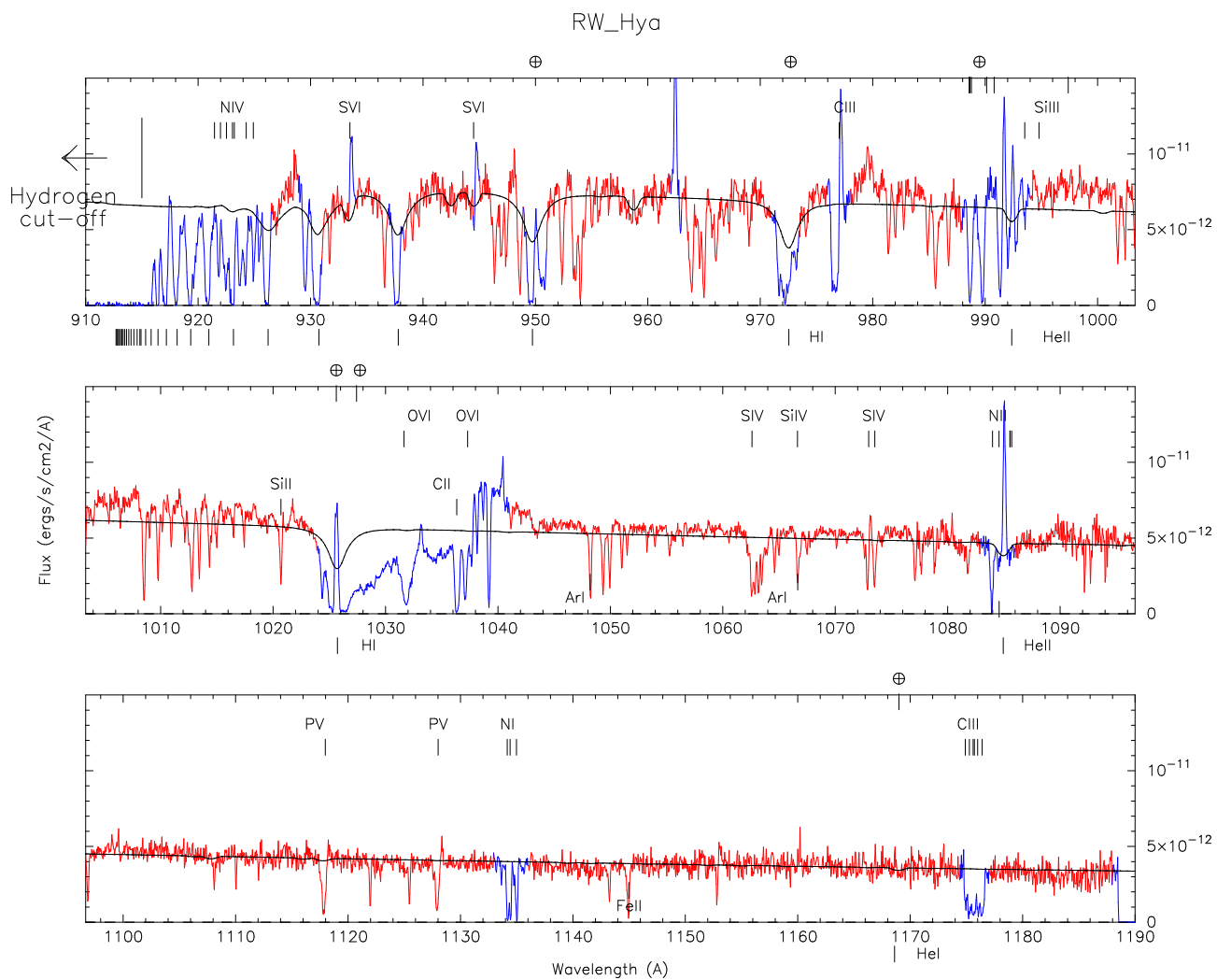}
\vspace{1.cm} 
\caption{A single temperature hot NLTE WD model fit for a low mass 
WD to the  dereddened FUSE spectrum of RW Hya with a best fit corresponding to 
a white dwarf with a surface temperature of 160,000K and 
$\log(g)=6.5$. Assuming a mass of $0.4M_{\odot}$ with a radius 
of 0.065$R_{\odot}$ yields 
a scale factor-derived distance of 811 pc, confirming that a low mass 
white dwarf with its larger radius, and a very high temperature gives 
a distance to RW Hya which is within the error bars of the original 
Schild et al. distance.}
\end{figure}

\begin{figure}[!ht]
\vspace{-5.cm} 
\plotone{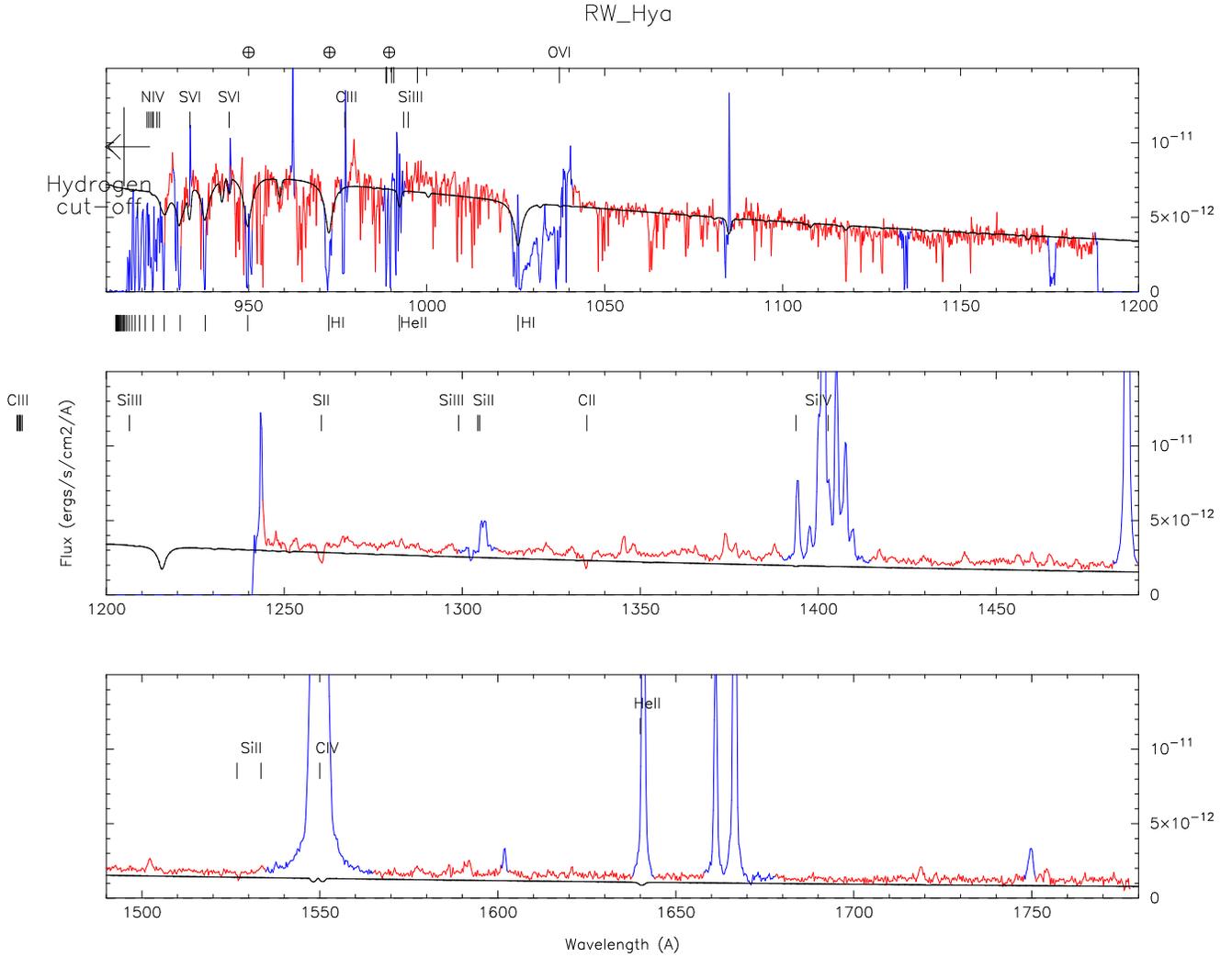}
\vspace{1.cm} 
\caption{
The combined  dereddened FUSE+ HST GHRS spectrum of RW Hya is modeld as in
the previous figure. 
A very hot and very small mass WD provides the required flux. The 
fit reveals an accreting white dwarf with 
$T_{wd}=160,000$K, $\log(g)=6.5$. The two spectra were obtained 
at different epochs and with different instruments and the HST spectrum
has a flux slightly lower than in the FUSE spectrum.  
}
\end{figure}

\begin{figure}[!ht]
\vspace{-5.cm} 
\plotone{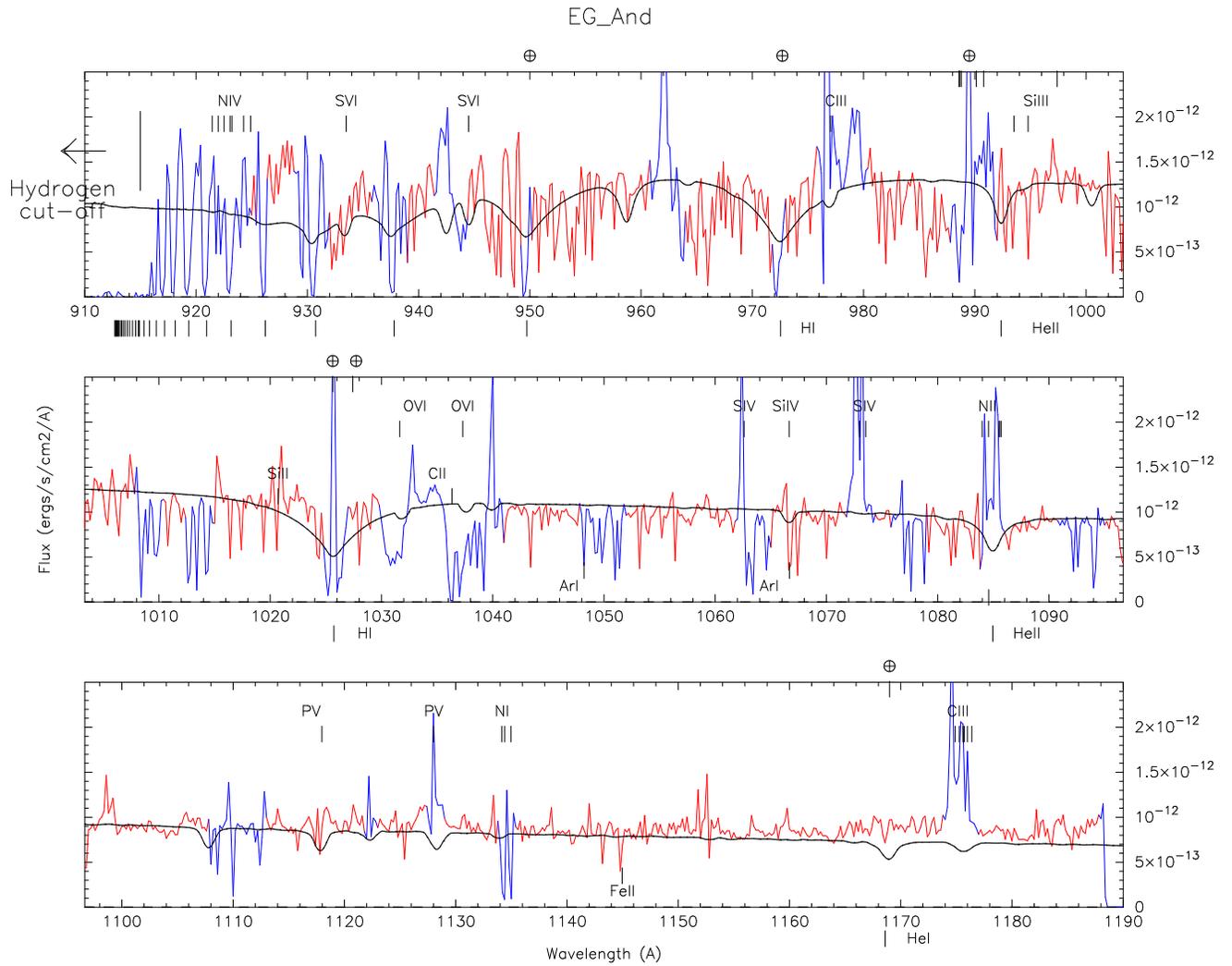}
\vspace{1.cm} 
\caption{The best-fitting model solution for the dereddened FUSE spectrum of EG And. 
The hot component is revealed to be a hot, bare white dwarf with 
a surface temperature of 90,000K and $\log(g)= 7.5$.}
\end{figure}

\begin{figure}[!ht]
\vspace{-5.cm} 
\plotone{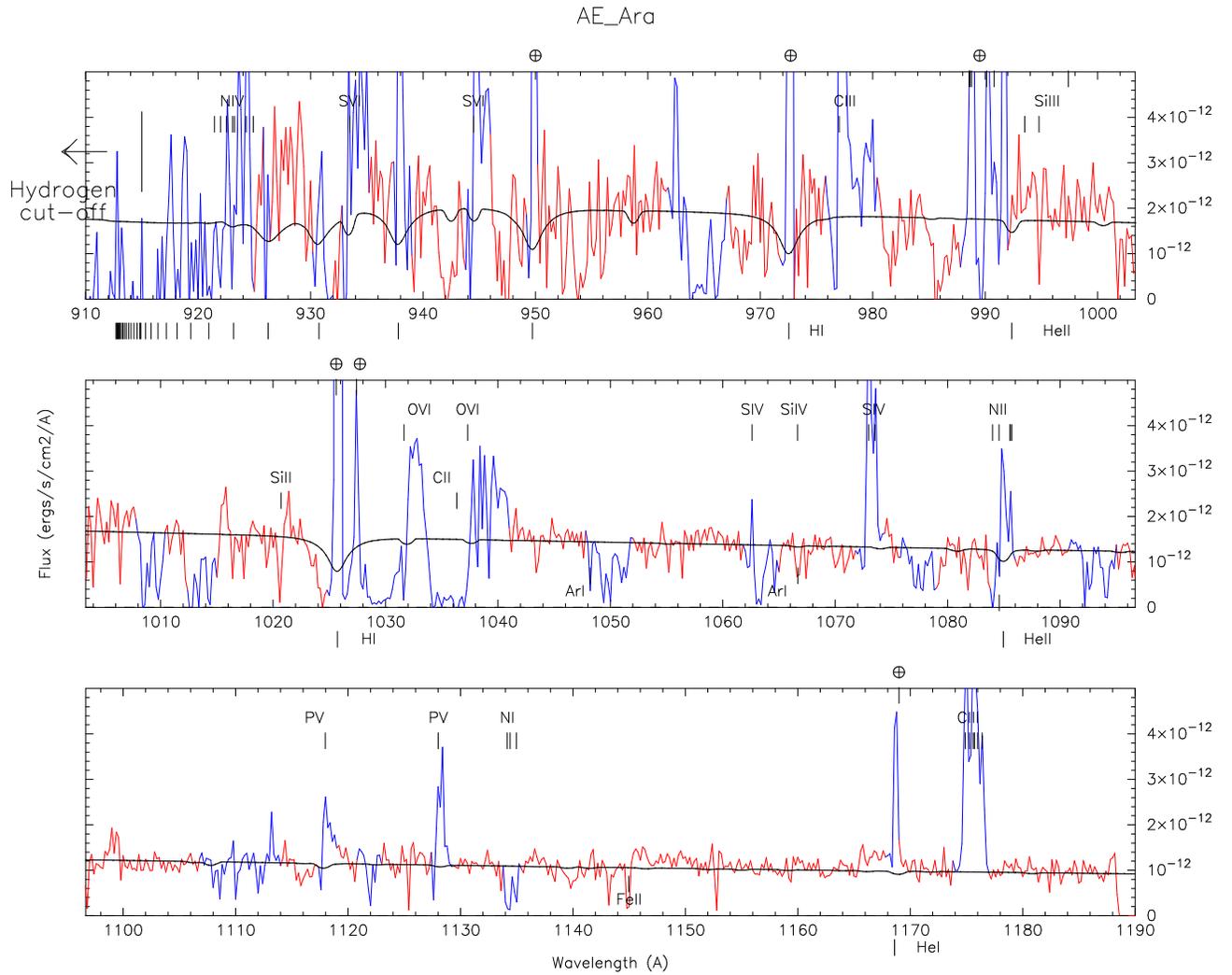}
\vspace{1.cm} 
\caption
{The dereddened FUSE spectrum of AE Ara has been fitted with a single WD model 
with $\log(g)= 6$ and $T_{\rm eff}=130,000$ K. 
} 
\end{figure}

\end{document}